\documentclass[aip,jmp,amsmath,amssymb,reprint]{revtex4-1}
\pdfoutput=1          
\usepackage{graphics} 
\usepackage{graphicx} 
\usepackage{dcolumn}  
\usepackage{bm}       
\usepackage{natbib}

\begin{document}

\title[Numerical testing of Simmons' equation by a transfer-matrix technique]{Numerical testing by a transfer-matrix technique of Simmons' equation for the local current density in metal-vacuum-metal junctions}

\author{Alexandre Mayer}
\email{alexandre.mayer@unamur.be}
\affiliation{Department of Physics, University of Namur, Rue de Bruxelles 61, 5000 Namur, Belgium}

\author{Marwan S. Mousa}
\affiliation{Department of Physics, Mu’tah University, Al-Karak 61710, Jordan}

\author{Mark J. Hagmann}
\affiliation{Dept. Electrical and Computer Engineering, University of Utah, Salt Lake City, Utah, USA}

\author{Richard G. Forbes}
\affiliation{Advanced Technology Institute, University of Surrey, Guildford GU2 7XH, United Kingdom}

\date{18 December 2018; accepted for publication in Jordan Journal of Physics}

\begin{abstract}
We test the consistency with which Simmons' model can predict the local current density obtained for flat metal-vacuum-metal junctions. The image potential energy used in Simmons' original papers had a missing factor of $1/2$. Besides this technical issue, Simmons' model relies on a mean-barrier approximation for electron transmission through the potential-energy barrier between the metals. In order to test Simmons' expression for the local current density when the correct image potential energy is included, we compare the results of this expression with those provided by a transfer-matrix technique. This technique is known to provide numerically exact solutions of Schrodinger's equation for this barrier model. We also consider the current densities provided by a numerical integration of the transmission probability obtained with the WKB approximation and Simmons' mean-barrier approximation. The comparison between these different models shows that Simmons' expression for the local current density actually provides results that are in good agreement with those provided by the transfer-matrix technique, for a range of conditions of practical interest.
We show that Simmons' model provides good results in the linear and field-emission regimes of current density versus voltage plots. It loses its applicability when the top of the potential-energy barrier drops below the Fermi level of the emitting metal.
\end{abstract}

\keywords{field electron emission, theory, metal-vacuum-metal junction, transmission probability, mean-barrier approximation, transfer-matrix technique}

\maketitle

\section{Introduction}

Analytical models are extremely useful for the study of field electron emission.
They provide indicative formulae for the emission current achieved
with given physical parameters. This enables quantitative understanding
of the role of these parameters. Analytical models also support the extraction of
useful information from experimental data. They certainly guide the development of
technologies. These analytical models depend however on a series of approximations,
typically the WKB (JWKB) approximation for the transmission of electrons through a
potential-energy barrier.\cite{Jeffreys_1925,Wentzel_1926,Kramers_1926,Brillouin_1926}
It is therefore natural to question the accuracy of these models.

The accuracy with which the Murphy-Good formulation of Fowler-Nordheim theory\cite{Fowler_Nordheim_1928,Murphy_Good_1956,Good_Muller_1956,Young_1959}
actually accounts for field electron emission from a flat metal surface was investigated in previous work.\cite{Forbes_JAP_2008,Forbes_JVSTB_2008,Mayer_2010_JPCM,Mayer_2010_JVSTB1,Mayer_2010_JVSTB2}
The approach adopted by Mayer consists in comparing the results of this analytical model with those provided by
a transfer-matrix technique.\cite{Mayer_2010_JPCM,Mayer_2010_JVSTB1,Mayer_2010_JVSTB2,Hagmann_1995}
This technique provides exact solutions of Schr\"odinger's equation for this field-emission process.
The comparison with the Murphy-Good expression $J_{\rm MG}=(\pi k_{\rm B}T/d)/\sin(\pi k_{\rm B}T/d)\times a t_{\rm F}^{-2}\Phi^{-1}F^2\exp[-bv_{\rm F}\Phi^{3/2}/F]$
for the current density obtained with an applied electrostatic field $F$, a work function $\Phi$ and a temperature $T$ revealed that the results of this analytical model are essentially correct, within a factor of the order 0.5-1. In the Murphy-Good expression, $a=1.541434\times 10^{-6}$ A eV V$^{-2}$, $b=6.830890$ eV$^{-3/2}$ V nm$^{-1}$,\cite{Forbes_JVSTB_2008} $k_{\rm B}$ is Boltzmann's constant, $t_{\rm F}$ and $v_{\rm F}$ are
particular values of well-known special mathematical functions that account for the image interaction,\cite{Good_Muller_1956,Forbes_2007} $d= \hbar e F / (2 t_{\rm F} \sqrt{2 m \Phi })$ with $e$ the
elementary positive charge and $m$ the electron mass. $\hbar$ is Planck's constant $h/2\pi$.
This study enabled the determination of a correction factor $\lambda^{\rm MG}$ to use with the Murphy-Good expression in order to get an exact result.\cite{Mayer_2010_JVSTB2}

The objective of the present work is to apply the same approach to the analytical model developed by Simmons
for the local current density through flat metal-vacuum-metal junctions.\cite{Simmons_1963_JAP1,Simmons_1963_JAP2,Simmons_1964_JAP1,Simmons_1964_JAP2,Matthews_2018}
Simmons' original model is widely cited in the literature.
It was however noted that the image potential energy used in the original papers
missed out a factor of $1/2$.\cite{Simmons_1964_JAP1,Miskovsky_1982}
An error in the current density obtained for a triangular barrier in the low-voltage range (Eq. 25 of Ref. \citenum{Simmons_1963_JAP1}) was also mentioned.\cite{Matthews_2018}
Besides these technical issues, Simmons' original model relies on a mean-barrier approximation for the transmission of electrons through the potential-energy barrier in the junction. It is natural to question this approximation and test the accuracy of the equation proposed by Simmons for the current density obtained in flat metal-vacuum-metal junctions when the correct image potential energy is included. We use for this purpose the transfer-matrix technique since it provides exact solutions for this barrier model. This work aims to provide a useful update and a numerical validation of Simmons' model.

This article is organized as follows. In Sec. \ref{section2}, we present the transfer-matrix technique that is used as
reference model for the quantum-mechanical simulation of metal-vacuum-metal junctions. In Sec. \ref{section3}, we present the main ideas of Simmons' theory. This presentation essentially focusses on the results that are discussed in this work.
In Sec. \ref{section4}, we compare the results of different models for the current density obtained in flat
metal-vacuum-metal junctions. We finally conclude this work in Sec. \ref{section5}.

\section{\label{section2}Modeling of metal-vacuum-metal junctions by a transfer-matrix technique}

\begin{figure}[ht]
 \begin{center}
  \begin{tabular}{c}
   \includegraphics[width=10cm]{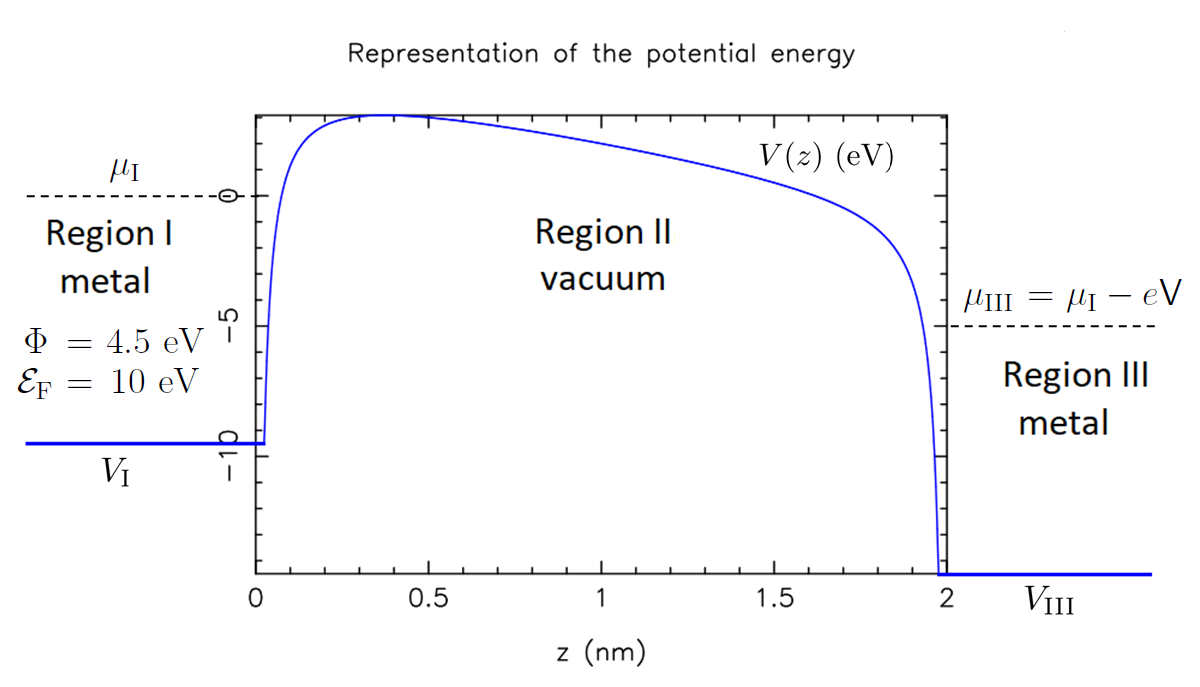}
  \end{tabular}
 \end{center}
 \caption{\label{figure1}Potential energy in a metal-vacuum-metal junction. A static voltage ${\sf V}$ of 5 V is applied.
  The gap spacing $D$ is 2 nm. We take for convenience the Fermi level $\mu_{\rm I}$ of the left-side metal
  as reference for the potential-energy values.}
\end{figure}

The metal-vacuum-metal junction considered in this work is represented in Fig. \ref{figure1}. For this particular example,
a static voltage ${\sf V}$ of 5 V is applied between the two metals. These metals have a Fermi energy ${\cal E}_{\rm F}$ of 10 eV and a common work function $\Phi$ of 4.5 eV. The gap spacing $D$ between the two metals is 2 nm. We refer by $\mu_{\rm I}$ to the Fermi level of the left-side metal (Region I). The Fermi level of the right-side metal (Region III)
is then given by $\mu_{\rm III}=\mu_{\rm I}-e{\sf V}$, where $e$ refers to the elementary positive charge.
For convenience when presenting Simmons' theory, we will use the Fermi level $\mu_{\rm I}$ of the left-side metal
as reference (zero value) for all potential-energy values discussed in this work. The total electron energy $E$ will also be defined with respect to $\mu_{\rm I}$. We will only consider positive
values for the applied voltage ${\sf V}$ so that the net electron current will always flow from the left to the right.
The potential energy in Region I and III is then given by $V_{\rm I}=\mu_{\rm I}-{\cal E}_{\rm F}$ and $V_{\rm III}=\mu_{\rm I}-e{\sf V}-{\cal E}_{\rm F}$. The potential energy in the vacuum gap ($0\leq z \leq D$) is given by $V(z)=\mu_{\rm I}+\Phi-eFz+V_{\rm image}(z)$, where $F={\sf V}/D$ is the magnitude of the electrostatic field induced by the voltage ${\sf V}$. $V_{\rm image}(z)$ refers to the image potential energy that applies to an electron
situated between two flat metallic surfaces (see Eq. \ref{Potential_Energy} in Sec. \ref{section3}). This vacuum region is also referred to as Region II.

In order to establish scattering solutions in cartesian coordinates, we assume that the wave functions are periodic along the lateral $x$ and $y$ directions (these directions are parallel to the flat surface of the two metals). We take a lateral periodicity $L$ of 10 nm for the wave functions (this value is sufficiently large to make our results independent of $L$). The boundary states in Region I and III are given respectively by
$\Psi_{i,j}^{\rm I,\pm}({\bf r},t) = e^{{\rm i} (k_{{\rm x},i} x + k_{{\rm y},j} y)}
 e^{\pm {\rm i} \sqrt{\frac{2m}{\hbar^{2}}(E-V_{\rm I})-k_{{\rm x},i}^{2}-k_{{\rm y},j}^{2}}  z}
 e^{-{\rm i} E t/\hbar}$ and $\Psi_{i,j}^{\rm III,\pm}({\bf r},t) = e^{{\rm i} (k_{{\rm x},i} x + k_{{\rm y},j} y)}
 \linebreak
 e^{\pm {\rm i} \sqrt{\frac{2m}{\hbar^{2}}(E-V_{\rm III})-k_{{\rm x},i}^{2}-k_{{\rm y},j}^{2}}  z}
 e^{-{\rm i} E t/\hbar}$, where ${\rm i}=\sqrt{-1}$ and the $\pm$ signs refer to the propagation direction
 of these boundary states relative to the $z$-axis. $E$ is the total electron energy. $k_{x,i}=i \frac{2\pi}{L}$ and $k_{y,j}=j \frac{2\pi}{L}$ are the lateral components of the wavevector ($i$ and $j$ are two integers also used to enumerate the boundary states). $E_{\rm z}=E-\frac{\hbar^2}{2m}(k_{{\rm x},i}^2+k_{{\rm y},j}^2)$ corresponds to the normal component of the electron energy.

By using a transfer-matrix technique, we can establish scattering solutions of Schr\"odinger's equation
$[\frac{\hbar^2}{2m}\Delta + V({\bf r})]\Psi({\bf r},t)={\rm i}\hbar \frac{\partial}{\partial t}\Psi({\bf r},t)$.
The idea consists in propagating the boundary states $\Psi_{i,j}^{{\rm III},\pm}$ of Region III across the vacuum gap (Region II).
Since the potential energy is independent of $x$ and $y$, there is no coupling between states associated with different
values of $i$ or $j$ and one can consider the propagation of these states separately. For the propagation of these states,
we assume that the potential energy in Region II varies in steps of width $\Delta z$ along the direction $z$. For each
integer $s$ ranging backwards from $D/\Delta z$ to 1, the potential energy is thus replaced by the constant value
$V_s=\frac{1}{2}[V((s-1).\Delta z)+V(s.\Delta z)]$. The solutions of Schr\"odinger's equation are then
(i) simple plane waves $A_s\ e^{i\sqrt{\frac{2m}{\hbar^2}(E_{\rm z}-V_s)}z}+B_s\ e^{-i\sqrt{\frac{2m}{\hbar^2}(E_{\rm z}-V_s)}z}$
when $E_{\rm z}=E-\frac{\hbar^2}{2m}(k_{{\rm x},i}^2+k_{{\rm y},j}^2)>V_s$,
(ii) real exponentials $A_s\ e^{-\sqrt{\frac{2m}{\hbar^2}(V_s-E_{\rm z})}z}+B_s\ e^{\sqrt{\frac{2m}{\hbar^2}(V_s-E_{\rm z})}z}$
when $E_{\rm z}<V_s$ or (iii) linear functions $A_s + B_s\ z$ when $E_{\rm z}=V_s$.
One can get arbitrarily close to the exact potential-energy barrier by letting $\Delta z \to 0$ (we used $\Delta
z$=0.0001 nm). The propagation of the states $\Psi_{i,j}^{{\rm III},\pm}$ across Region II is
then achieved by matching continuity conditions for the wave function $\Psi$ and its derivative $\frac{d\Psi}{dz}$
at the boundaries of each step $\Delta z$, when going backwards from $z=D$ to $z=0$.\cite{Mayer_2010_JPCM} The layer-addition algorithm presented in a previous work should be used to prevent numerical instabilities.\cite{Mayer_PRE_1999} The solutions finally obtained
for $z=0$ are expressed as linear combinations of the boundary states $\Psi_{i,j}^{\rm I,\pm}$ in Region I.

This procedure leads to the following set of solutions :
\begin{eqnarray}
 \hat\Psi_{i,j}^{+} &\stackrel{z\leq 0}{=}& T_{i,j}^{++} \Psi_{i,j}^{\rm I,+} + T_{i,j}^{-+} \Psi_{i,j}^{\rm I,-} \stackrel{z\geq D }{=} \Psi_{i,j}^{\rm III,+},\label{Psi_T_1} \\
 \hat\Psi_{i,j}^{-} &\stackrel{z\leq 0}{=}& T_{i,j}^{+-} \Psi_{i,j}^{\rm I,+} + T_{i,j}^{--} \Psi_{i,j}^{\rm I,-} \stackrel{z\geq D }{=} \Psi_{i,j}^{\rm III,-},\label{Psi_T_2}
\end{eqnarray}
where the complex numbers $T_{i,j}^{\pm\pm}$ correspond to the coefficients of these solutions in Region I.

We can then take linear combinations of these solutions
in order to establish scattering solutions that correspond to single incident states $\Psi_{i,j}^{\rm I,+}$ in Region I or
$\Psi_{i,j}^{\rm III,-}$ in Region III. These solutions will have the form
\begin{eqnarray}
 \Psi_{i,j}^{+} &\stackrel{z\leq 0}{=}& \Psi_{i,j}^{\rm I,+} + S_{i,j}^{-+} \Psi_{i,j}^{\rm I,-} \stackrel{z\geq D }{=} S_{i,j}^{++} \Psi_{i,j}^{\rm III,+},\label{Psi_S_1}\\
 \Psi_{i,j}^{-} &\stackrel{z\leq 0}{=}& S_{i,j}^{--} \Psi_{i,j}^{\rm I,-} \stackrel{z\geq D }{=} \Psi_{i,j}^{\rm III,-}+ S_{i,j}^{+-} \Psi_{i,j}^{\rm III,+},\label{Psi_S_2}
\end{eqnarray}
where the complex numbers $S_{i,j}^{++}$ and $S_{i,j}^{-+}$ provide respectively the coefficients of the transmitted
and reflected states for an incident state $\Psi_{i,j}^{\rm I,+}$ in Region I. The complex numbers $S_{i,j}^{--}$ and $S_{i,j}^{+-}$ provide respectively the coefficients of the transmitted and reflected states for an incident state $\Psi_{i,j}^{\rm III,-}$ in Region III. These coefficients are given by $S_{i,j}^{++}=[T_{i,j}^{++}]^{-1}$, $S_{i,j}^{-+}=T_{i,j}^{-+} [T_{i,j}^{++}]^{-1}$, $S_{i,j}^{--}= T_{i,j}^{--} - T_{i,j}^{-+} [T_{i,j}^{++}]^{-1} T_{i,j}^{+-}$ and $S_{i,j}^{+-}= - [T_{i,j}^{++}]^{-1} T_{i,j}^{+-}$.\cite{Footnote_2}

These scattering solutions are finally used to compute the local current density
$J$ that flows from Region I to Region III. The idea consists in integrating the contribution of each incident
state $\Psi_{i,j}^{\rm I,+}$ in Region I (this provides the current-density contribution moving to the right) as well as the contribution of each incident state $\Psi_{i,j}^{\rm III,-}$ in Region III (this provides the current-density contribution moving to the left). The net value of the current density is given by the difference between these two contributions. The detailed expression for the current density $J$ has been established in previous work.\cite{Mayer_PRB_1997,Mayer_PRB_2008,Mayer_JVSTB_2012} It is given formally by
\begin{eqnarray}
 J_{\rm TM} &=& \frac{1}{L^2} \frac{2e}{h} \int_{V_{\rm I}}^{\infty} \sum_{i,j} f_{\rm I}(E) \frac{v_{{\rm III},(i,j)}}{v_{{\rm I},(i,j)}} |S^{++}_{i,j}|^2 dE \nonumber\\
  &-& \frac{1}{L^2} \frac{2e}{h} \int_{V_{\rm III}}^{\infty} \sum_{i,j} f_{\rm III}(E) \frac{v_{{\rm I},(i,j)}}{v_{{\rm III},(i,j)}} |S^{--}_{i,j}|^2 dE ,
 \label{J_TM}
\end{eqnarray}
where the summations are restricted to solutions that are propagative both in Region I and Region III.
This requires $E_{\rm z}=E-\frac{\hbar^2}{2m}(k_{{\rm x},i}^2+k_{{\rm y},j}^2)>\max(V_{\rm I},V_{\rm III})$.
$v_{{\rm I},(i,j)}=\frac{\hbar}{m}\sqrt{\frac{2m}{\hbar^2}(E_{\rm z}-V_{\rm I})}$
and $v_{{\rm III},(i,j)}=\frac{\hbar}{m}\sqrt{\frac{2m}{\hbar^2}(E_{\rm z}-V_{\rm III})}$
represent the normal component of the electron velocity in Region I and III.
$\frac{v_{{\rm III},(i,j)}}{v_{{\rm I},(i,j)}} |S^{++}_{i,j}|^2$ and
$\frac{v_{{\rm I},(i,j)}}{v_{{\rm III},(i,j)}} |S^{--}_{i,j}|^2$ both represent the
transmission probability ${\cal D}_{\rm TM}$ of the potential-energy barrier in Region II,
at the normal energy $E_{\rm z}$.
$f_{\rm I}(E)=1/\{1+\exp[(E-\mu_{\rm I})/k_{\rm B}T]\}$
and $f_{\rm III}(E)=1/\{1+\exp[(E-\mu_{\rm III})/k_{\rm B}T]\}$ finally
refer to the Fermi distributions in Region I and III.\cite{Footnote_3}

One can show mathematically that Eq. \ref{J_TM}, with $L\gg 1$, is equivalent to
\begin{eqnarray}
 J_{\rm TM} &=& \int_{\max(V_{\rm I},V_{\rm III})}^\infty \Delta {\cal N}(E_{\rm z}) {\cal D}_{\rm TM}\left(E_{\rm z}\right)dE_{\rm z},\label{J_TM_Ez}
\end{eqnarray}
where the integration is over the normal energy $E_{\rm z}$ instead of the total energy $E$.
${\cal D}_{\rm TM}\left(E_{\rm z}\right) = \frac{v_{{\rm III},(i,j)}}{v_{{\rm I},(i,j)}} |S^{++}_{i,j}|^2 =
\frac{v_{{\rm I},(i,j)}}{v_{{\rm III},(i,j)}} |S^{--}_{i,j}|^2$ is the transmission probability
of the potential-energy barrier at the normal energy $E_{\rm z}$. $\Delta {\cal N}(E_{\rm z}) = {\cal N}_{\rm I}(E_{\rm z})-{\cal N}_{\rm III}(E_{\rm z})$, with ${\cal N}_{\rm I}(E_{\rm z}) = \frac{4\pi m e}{h^3} k_{\rm B}T
\linebreak
\ln\left[1+\exp\left(-\frac{E_{\rm z}-\mu_{\rm I}}{k_{\rm B}T}\right)\right]$ and ${\cal N}_{\rm III}(E_{\rm z})=\frac{4\pi m e}{h^3} k_{\rm B}T
\ln\left[1+\exp\left(-\frac{E_{\rm z}-\mu_{\rm I}+e{\sf V}}{k_{\rm B}T}\right)\right]$
the incident normal-energy distributions of the two metals. This expression
of the local current density is more standard in the field emission community.

For the integration over $E$ in Eq. \ref{J_TM} or $E_{\rm z}$ in Eq. \ref{J_TM_Ez}, we use a step
$\Delta E$ of 0.01 eV. It was checked that Eq. \ref{J_TM} and \ref{J_TM_Ez} provide identical results.
A room temperature $T$ of 300 K is assumed in this work.

\section{\label{section3}Simmons' model for the current density in flat metal-vacuum-metal junctions}

We present now the main ideas of Simmons' model for the local current density through a flat metal-vacuum-metal junction
(see Fig. \ref{figure1}). This presentation focuses on the results that are actually required for a
comparison with the transfer-matrix results. We keep for consistency the notations introduced in the previous section.

\subsection{Potential-energy barrier}

The potential energy in the vacuum gap ($0\leq z \leq D$) is given by\cite{Simmons_1963_JAP1}
\begin{eqnarray}
 V(z)&=&\mu_{\rm I}+\Phi-eFz 
      -\frac{1}{2}\frac{e^2}{4\pi\epsilon_0}\left[\frac{1}{2z}+\sum_{n=1}^\infty \left(\frac{nD}{(nD)^2-z^2}-\frac{1}{nD}\right) \right],
 \label{Potential_Energy}
\end{eqnarray}
where the last term of Eq. \ref{Potential_Energy}
accounts for the image potential energy $V_{\rm image}(z)$ that applies to an electron situated
between two flat metallic surfaces.\cite{Footnote_1}
In Simmons' original papers,\cite{Simmons_1963_JAP1,Simmons_1963_JAP2}
there is a factor $1/2$ missing in the image potential energy. This factor $1/2$, which is included for correction in
Eq. \ref{Potential_Energy}, comes from the self-interaction character of the image potential energy (the image charges
follow automatically the displacement of the electron and work must actually only be done on the electron). This
technical error was mentioned later by Simmons.\cite{Simmons_1964_JAP1} It was also pointed out in a paper by
Miskovsky et al.\cite{Miskovsky_1982}

In order to derive analytical expressions for the local current density, Simmons introduces a useful
approximation for the image potential energy : $V_{\rm image}(z)\simeq -1.15\lambda\frac{D^2}{z(D-z)}$.\cite{Simmons_1963_JAP1}
The potential energy in the vacuum gap can then be approximated by
\begin{eqnarray}
 V(z)=\mu_{\rm I}+\Phi-eFz-1.15\lambda\frac{D^2}{z(D-z)},
 \label{Potential_Energy_Approx}
\end{eqnarray}
where $\lambda=\frac{e^2}{16\pi\epsilon_0}\frac{\ln 2}{D}$. We provide here a corrected
expression for $\lambda$; this includes the missing factor $1/2$.

\subsection{Mean-barrier approximation for the transmission probability}

With $E_{\rm z}=E-\frac{\hbar^2}{2m}(k_{\rm x}^2+k_{\rm y}^2)$ the normal component of the energy, the probability for
an electron to cross the potential-energy barrier in Region II is given, within the simple WKB approximation,\cite{Jeffreys_1925,Wentzel_1926,Kramers_1926,Brillouin_1926} by
\begin{eqnarray}
 {\cal D}_{\rm WKB} &=& \exp\left\{ -\frac{2\sqrt{2m}}{\hbar}\int_{z_1}^{z_2}[V(z)-E_{\rm z}]^{1/2} dz \right\},
 \label{T_WKB}
\end{eqnarray}
where $z_1$ and $z_2$ are the classical turning points of the barrier at the normal energy $E_{\rm z}$
(i.e., the solutions of $V(z_1)=V(z_2)=E_{\rm z}$ with $z_1\leq z_2$). Simmons then replaces $V(z)$ by $V(z)=\mu_{\rm I}+\phi(z)$,
where $\phi(z)=\Phi-eFz+V_{\rm image}(z)$ represents the difference between $V(z)$ and the Fermi level $\mu_{\rm I}$ of the left-side metal (this is the metal that actually emits electrons for a positive voltage). He finally proposes a mean-barrier approximation for the transmission probability\cite{Simmons_1963_JAP1}\,:
\begin{eqnarray}
 {\cal D}_{\rm Sim} &=& \exp\left\{ -\frac{2\sqrt{2m}}{\hbar}\ \beta\ \Delta z\ [\overline\phi-(E_{\rm z}-\mu_{\rm I})]^{1/2} \right\},
 \label{T_Simmons}
\end{eqnarray}
where $\Delta z=z_2-z_1$ represents here the width of the barrier at the Fermi level of the left-side metal
(i.e., for $E_{\rm z}=\mu_{\rm I}$).
$\overline\phi=\frac{1}{z_2-z_1}\int_{z_1}^{z_2}\phi(z) dz$ represents the mean barrier height
above the Fermi level of the left-side metal.
$\beta$ is a correction factor related to the mean-square deviation of $\phi(z)$ with respect to $\overline\phi$.\cite{Simmons_1963_JAP1}
For the barrier shown in Eq. \ref{Potential_Energy} (image potential energy included), Simmons recommends using $\beta=1$.
The mathematical justification of Eq. \ref{T_Simmons} can be found in the Appendix of Ref. \citenum{Simmons_1963_JAP1}.

\subsection{Analytical expression for the local current density}

In his original paper,\cite{Simmons_1963_JAP1} Simmons proposes a general formula for the net local current density $J$
that flows between the two metals of the junction (see Eq. 20 of Ref. \citenum{Simmons_1963_JAP1}). The idea consists
in integrating the contribution to the current density of each incident state in the two metals
(the transmission of these states through the potential-energy barrier is evaluated with Eq. \ref{T_Simmons}).
Different analytical approximations were introduced by Simmons to achieve this result
(in particular, in Eqs 15, 16 and 18 that lead to Eq. 20 of Ref. \citenum{Simmons_1963_JAP1};
they require $\frac{2\sqrt{2m}}{\hbar}\beta\Delta z (\overline{\phi}+e{\sf V})^{1/2} \gg 1$).
The temperature-dependence of the current density was established in Ref. \citenum{Simmons_1964_JAP2}.
The final expression, which accounts for the temperature, is given by
\begin{eqnarray}
 J_{\rm Sim} &=& J_0\times \frac{\pi B k_{\rm B} T}{\sin(\pi B k_{\rm B} T)}\times
 \left\{ \overline{\phi}\ \exp\left(- A\ \overline{\phi}^{1/2} \right)\right.\nonumber\\
                 & & \left. -(\overline{\phi}+e{\sf V})\ \exp\left(-A\ (\overline{\phi}+e{\sf V})^{1/2} \right)\right\},
 \label{J_Simmons}
\end{eqnarray}
where $J_0=\frac{e}{\hbar(2\pi\beta \Delta z)^2}$, $A=\frac{2\sqrt{2m}}{\hbar}\beta\Delta z$ and
$B = \frac{A}{2\overline{\phi}^{1/2}}$. The term $J_0\ \overline{\phi}\ \exp\left(- A\ \overline{\phi}^{1/2} \right)$
accounts for the current moving to the right.
The term $J_0\ (\overline{\phi}+e{\sf V})\ \exp\left(-A\ (\overline{\phi}+e{\sf V})^{1/2} \right)$ accounts for the
current moving to the left. The temperature-dependence is contained in the factor
$\frac{\pi B k_{\rm B} T}{\sin(\pi B k_{\rm B} T)}$.\cite{Simmons_1964_JAP2,Hrach_1968}.
As mentioned previously, a temperature $T$ of 300 K is considered in this work.

For a potential-energy barrier approximated by Eq. \ref{Potential_Energy_Approx}, Simmons provides an approximation for
the classical turning points at the Fermi level of the left-side metal.\cite{Simmons_1963_JAP1} If $e{\sf V}<\Phi$, with $\Phi$ the local work function, these turning points are given by
\begin{eqnarray}
 \left\{
  \begin{array}{l}
   z_1 = 1.2\lambda D/\Phi \cr
   z_2 = D [1-9.2\lambda/(3\Phi+4\lambda-2e{\sf V})]+z_1\cr
  \end{array}
 \right..
 \label{Simmons_z1z2_1}
\end{eqnarray}
Otherwise, if $e{\sf V}\geq \Phi$, they are given by
\begin{eqnarray}
 \left\{
  \begin{array}{l}
   z_1 = 1.2\lambda D/\Phi \cr
   z_2 = (\Phi-5.6\lambda)(D/e{\sf V}) \cr
  \end{array}
 \right..
 \label{Simmons_z1z2_2}
\end{eqnarray}
These expressions are calculated with the corrected factor $\lambda=\frac{e^2}{16\pi\epsilon_0}\frac{\ln 2}{D}$.
We can then compute the width $\Delta z=z_2-z_1$ of the barrier at the Fermi level of the left-side metal
as well as the mean barrier height $\overline{\phi}$ above this Fermi level
($\overline{\phi}$ represents the mean barrier height, over the range $\Delta z$, experienced by an
electron tunneling with a normal energy equal to the left-side Fermi level).\cite{Simmons_1963_JAP1}
The result is given by
\begin{eqnarray}
 \overline{\phi} &=& \Phi-\frac{e{\sf V}(z_1+z_2)}{2D}-\frac{1.15\lambda D}{z_2-z_1}\ln\left[\frac{z_2(D-z_1)}{z_1(D-z_2)}\right].
 \label{Simmons_Vmoy}
\end{eqnarray}
With Simmons' recommendation to use $\beta=1$, we can compute each quantity in Eq. \ref{J_Simmons}. This is the equation
we want to test numerically by comparing its predictions with the results of the transfer-matrix technique.
$J_{\rm Sim}$ depends on the mean-barrier approximation of the transmission probability (Eq. \ref{T_Simmons}),
on the analytical approximations introduced by Simmons to establish Eq. \ref{J_Simmons} and on Eqs \ref{Simmons_z1z2_1},
\ref{Simmons_z1z2_2} and \ref{Simmons_Vmoy} for $\Delta z=z_2-z_1$ and $\overline{\phi}$.

\subsection{Numerical expressions for the local current density}

It is actually possible to integrate numerically the transmission probability ${\cal D}_{\rm Sim}$ provided by Eq. \ref{T_Simmons}.
By analogy with the current density $J_{\rm TM}$ provided by the transfer-matrix formalism, the current density obtained
by the numerical integration of ${\cal D}_{\rm Sim}$ will be given by
\begin{widetext}
\begin{eqnarray}
 J_{\rm Sim-num}
 &=& \frac{1}{L^2} \frac{2e}{h} \int_{V_{\rm I}}^{\infty}
 \sum_{i,j} f_{\rm I}(E) {\cal D}_{\rm Sim}\left(E-\frac{\hbar^2}{2m}(k_{{\rm x},i}^2+k_{{\rm y},j}^2)\right)dE\nonumber\\
   &-& \frac{1}{L^2} \frac{2e}{h} \int_{V_{\rm III}}^{\infty}
 \sum_{i,j} f_{\rm III}(E) {\cal D}_{\rm Sim}\left(E-\frac{\hbar^2}{2m}(k_{{\rm x},i}^2+k_{{\rm y},j}^2)\right)dE\label{J_Simmons_Numeric}\\
 &=& \int_{\max(V_{\rm I},V_{\rm III})}^\infty \Delta {\cal N}(E_{\rm z}) {\cal D}_{\rm Sim}\left(E_{\rm z}\right)dE_{\rm z}\label{J_Simmons_Numeric_Ez}
\end{eqnarray}
\end{widetext}
in the standard formulation. ${\cal D}_{\rm Sim}$ is obtained here by a
numerical evaluation of Eq. \ref{T_Simmons}
($\Delta z=z_2-z_1$ and $\overline{\phi}$ are evaluated on the exact barrier given in Eq. \ref{Potential_Energy}).
The comparison of $J_{\rm Sim-num}$ with the results of Eq. \ref{J_Simmons} will validate the approximations that lead to this analytical expression.

It will also be interesting to consider the current density obtained by a numerical integration of the transmission probability provided by the simple WKB approximation (Eq. \ref{T_WKB}). The result will be given by
\begin{widetext}
\begin{eqnarray}
 J_{\rm WKB}
 &=& \frac{1}{L^2} \frac{2e}{h} \int_{V_{\rm I}}^{\infty}
 \sum_{i,j} f_{\rm I}(E) {\cal D}_{\rm WKB}\left(E-\frac{\hbar^2}{2m}(k_{{\rm x},i}^2+k_{{\rm y},j}^2)\right)dE\nonumber\\
   &-& \frac{1}{L^2} \frac{2e}{h} \int_{V_{\rm III}}^{\infty}
 \sum_{i,j} f_{\rm III}(E) {\cal D}_{\rm WKB}\left(E-\frac{\hbar^2}{2m}(k_{{\rm x},i}^2+k_{{\rm y},j}^2)\right)dE\label{J_WKB}\\
 &=& \int_{\max(V_{\rm I},V_{\rm III})}^\infty \Delta {\cal N}(E_{\rm z}) {\cal D}_{\rm WKB}\left(E_{\rm z}\right)dE_{\rm z}\label{J_WKB_Ez}
\end{eqnarray}
\end{widetext}
in the standard formulation. $J_{\rm WKB}$ will enable a useful comparison with Simmons' theory given the fact that the transmission probability used by Simmons is actually an approximation of the WKB expression.

\section{\label{section4}Comparison between different models for the local current density}

We can compare at this point the local current densities provided by the transfer-matrix technique ($J_{\rm TM}$
by Eq. \ref{J_TM} or \ref{J_TM_Ez}), Simmons' analytical expression ($J_{\rm Sim}$ by Eq. \ref{J_Simmons}), a numerical integration of Simmons' formula for the transmission probability ($J_{\rm Sim-num}$ by Eq. \ref{J_Simmons_Numeric_Ez}) and a numerical integration of the transmission probability provided by the WKB approximation ($J_{\rm WKB}$ by Eq. \ref{J_WKB_Ez}).

In order to understand the different regimes that appear in typical $J$-{\sf V} plots, we will start by showing the $dJ/dE$ distributions obtained for a few representative cases. This will illustrate the "linear regime" and the "field-emission regime" that are indeed appropriately described by Simmons' equation \ref{J_Simmons}.
In the "linear regime", the difference $\mu_{\rm I}-\mu_{\rm III}$ between the Fermi level of the two metals is smaller than the width of the total-energy distribution of the right-flowing and left-flowing contributions to the current. These two contributions tend to cancel out except in an energy window of the order of $\mu_{\rm I}-\mu_{\rm III}$, which is equal to $e{\sf V}$.
In the "field-emission regime", the Fermi level $\mu_{\rm III}$ of the right metal is sufficiently far below $\mu_{\rm I}$ to make the contribution of the left-flowing current negligible. The diode current is essentially determined by the right-flowing current, which increases rapidly with {\sf V}.
The "flyover regime" will be beyond the predictive capacities of Simmons' theory.
In this regime, the top $V_{\rm top}$ of the potential-energy barrier drops below
$\mu_{\rm I}$ so that electrons at the Fermi level of the left metal can fly over the top of this barrier, provided $E_{\rm z}=E-\frac{\hbar^2}{2m}(k_{\rm x}^2+k_{\rm y}^2)>V_{\rm top}$.

\begin{figure}[h]
 \begin{center}
  \begin{tabular}{c}
   \includegraphics[height=8.5cm,angle=-90]{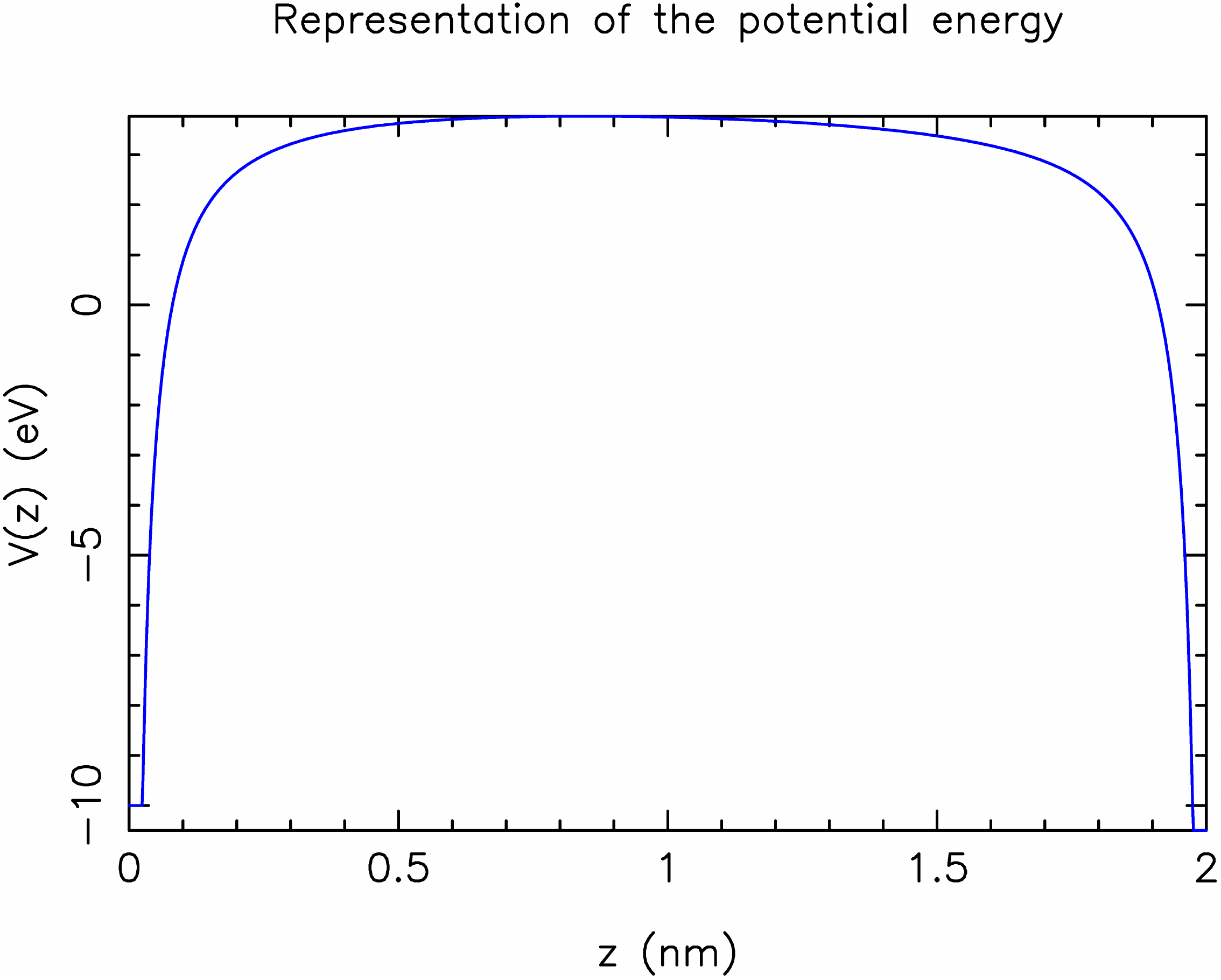} \cr
   \includegraphics[height=8.5cm,angle=-90]{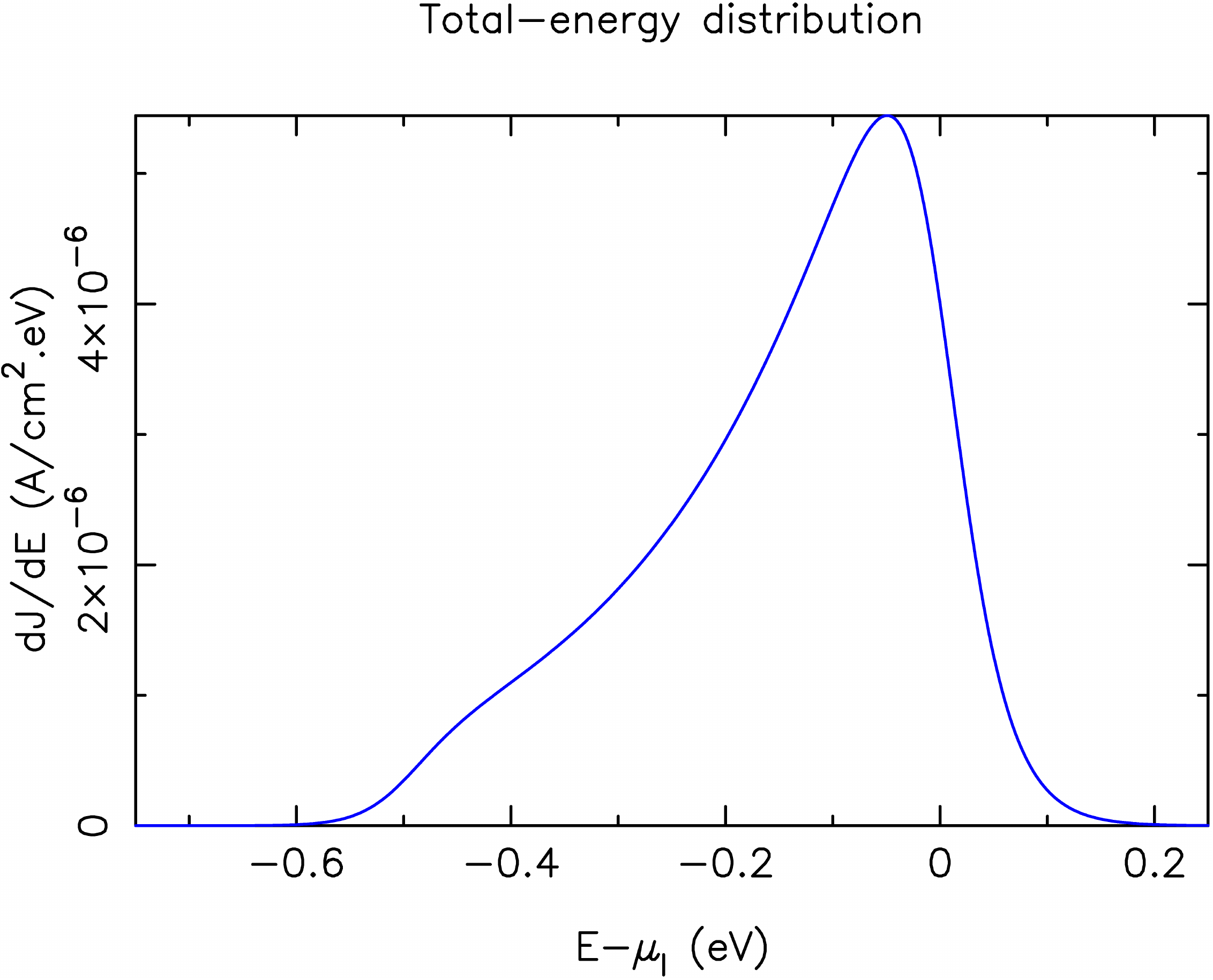} \cr
  \end{tabular}
 \end{center}
 \caption{\label{figure2}Potential energy $V(z)$ (top) and total-energy distribution of the current density $dJ/dE$ (bottom) for an applied voltage ${\sf V}$ of 0.5 V. $dJ/dE$ is calculated by the transfer-matrix technique. We take for convenience the Fermi level $\mu_{\rm I}$ of the left-side metal as reference for the potential-energy values.}
\end{figure}

\begin{figure}[h]
 \begin{center}
  \begin{tabular}{c}
   \includegraphics[height=8.5cm,angle=-90]{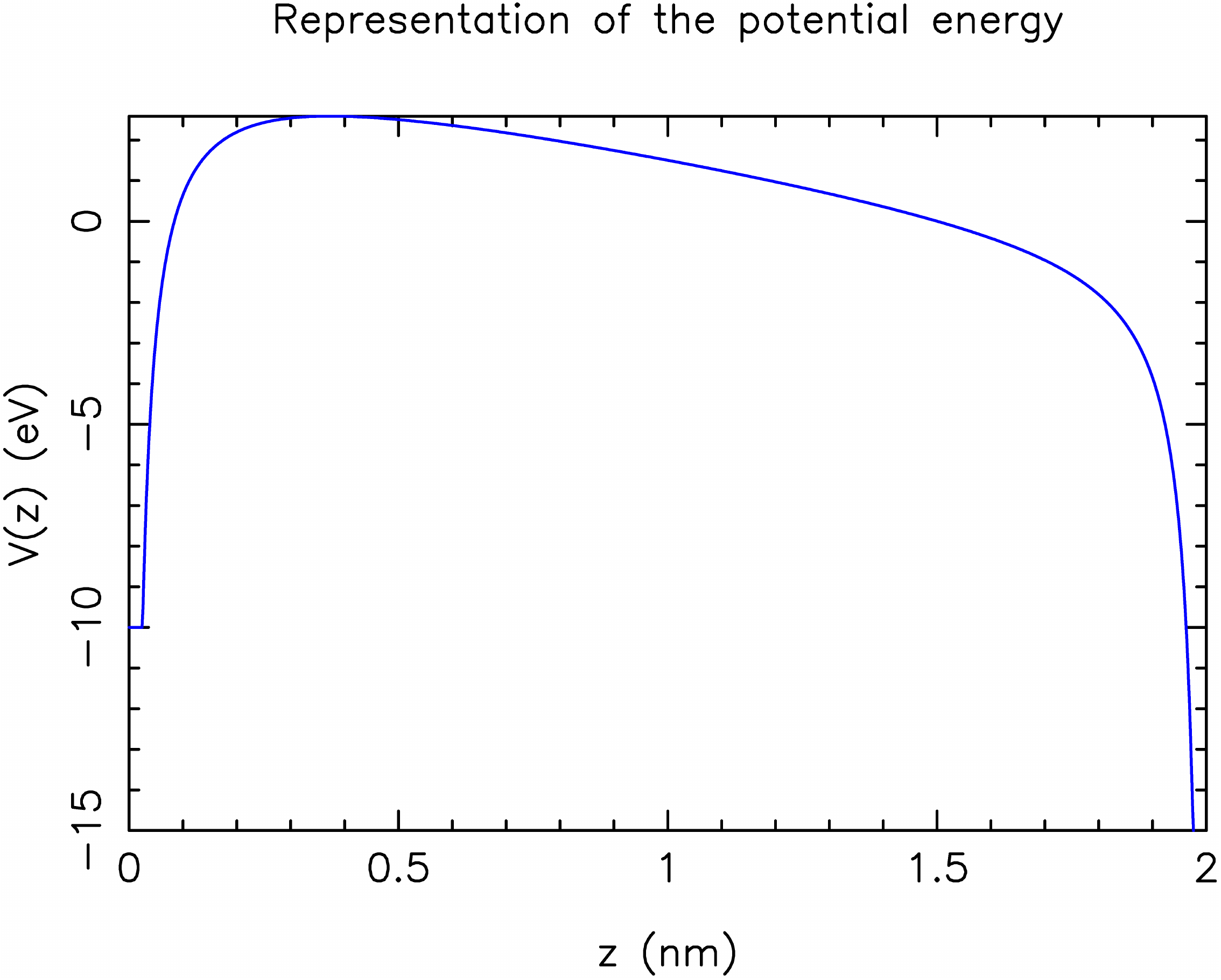} \cr
   \includegraphics[height=8.5cm,angle=-90]{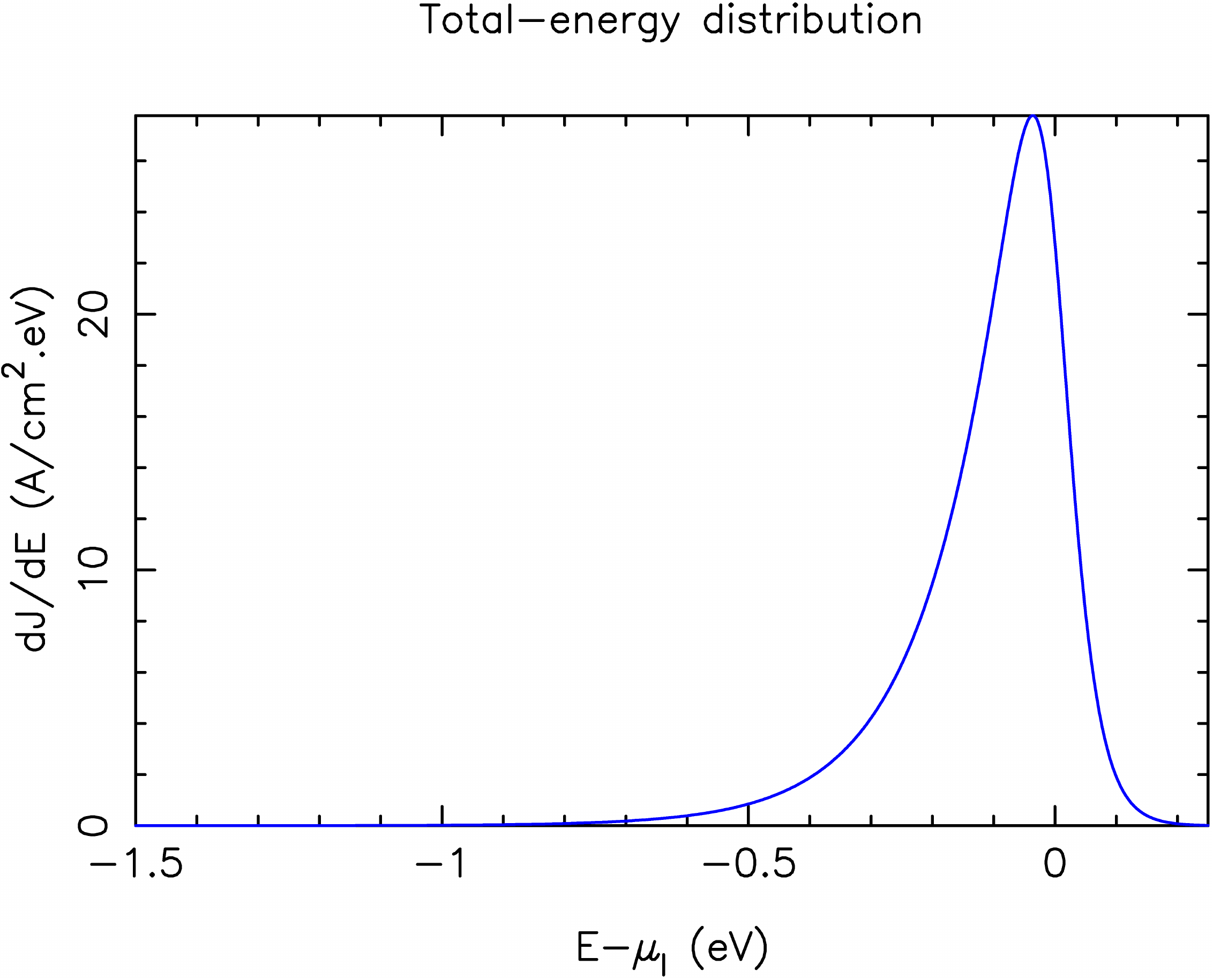} \cr
  \end{tabular}
 \end{center}
 \caption{\label{figure3}Potential energy $V(z)$ (top) and total-energy distribution of the current density $dJ/dE$ (bottom) for an applied voltage ${\sf V}$ of 5 V. $dJ/dE$ is calculated by the transfer-matrix technique. We take for convenience the Fermi level $\mu_{\rm I}$ of the left-side metal as reference for the potential-energy values.}
\end{figure}

\begin{figure}[h]
 \begin{center}
  \begin{tabular}{c}
   \includegraphics[height=8.5cm,angle=-90]{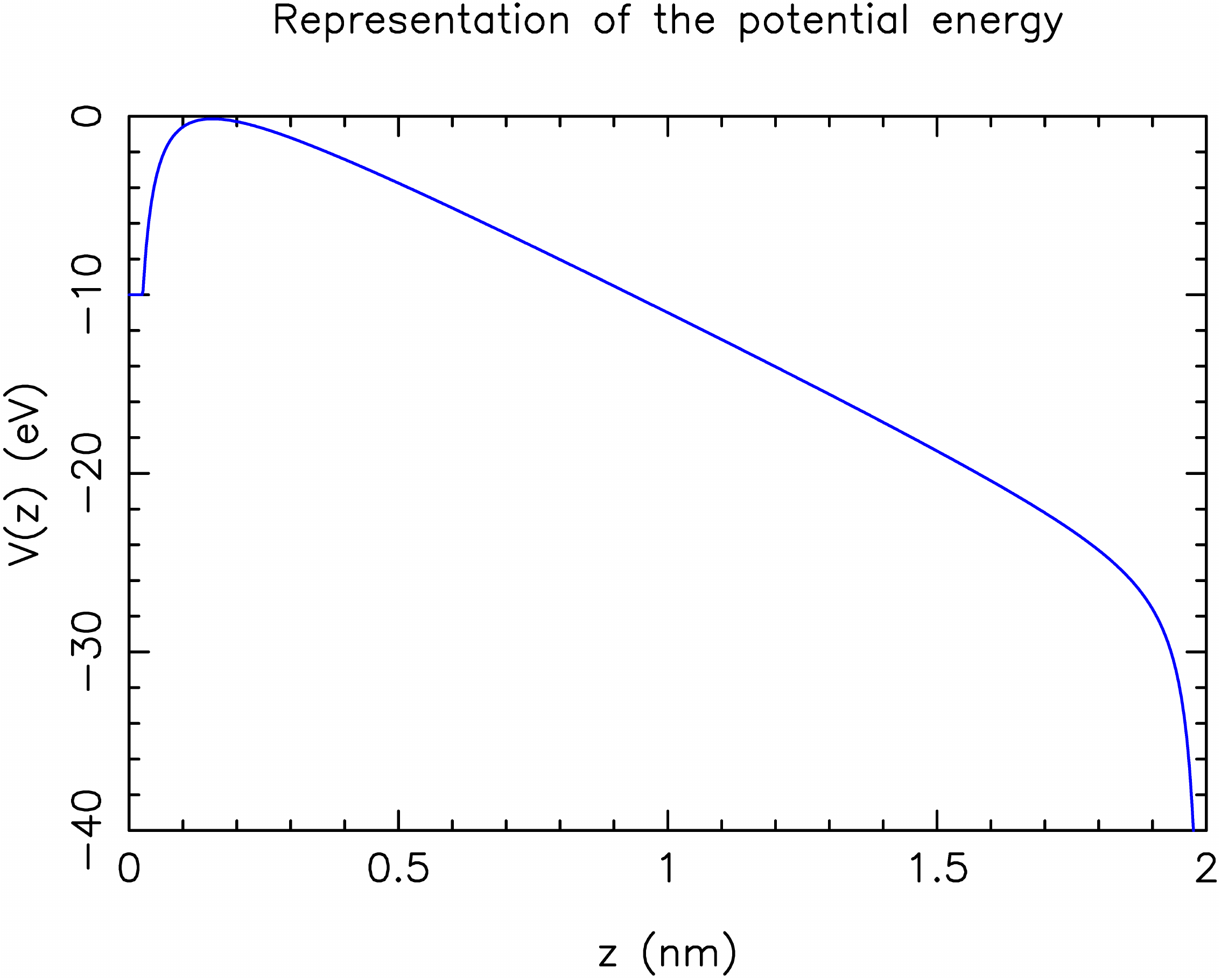} \cr
   \includegraphics[height=8.5cm,angle=-90]{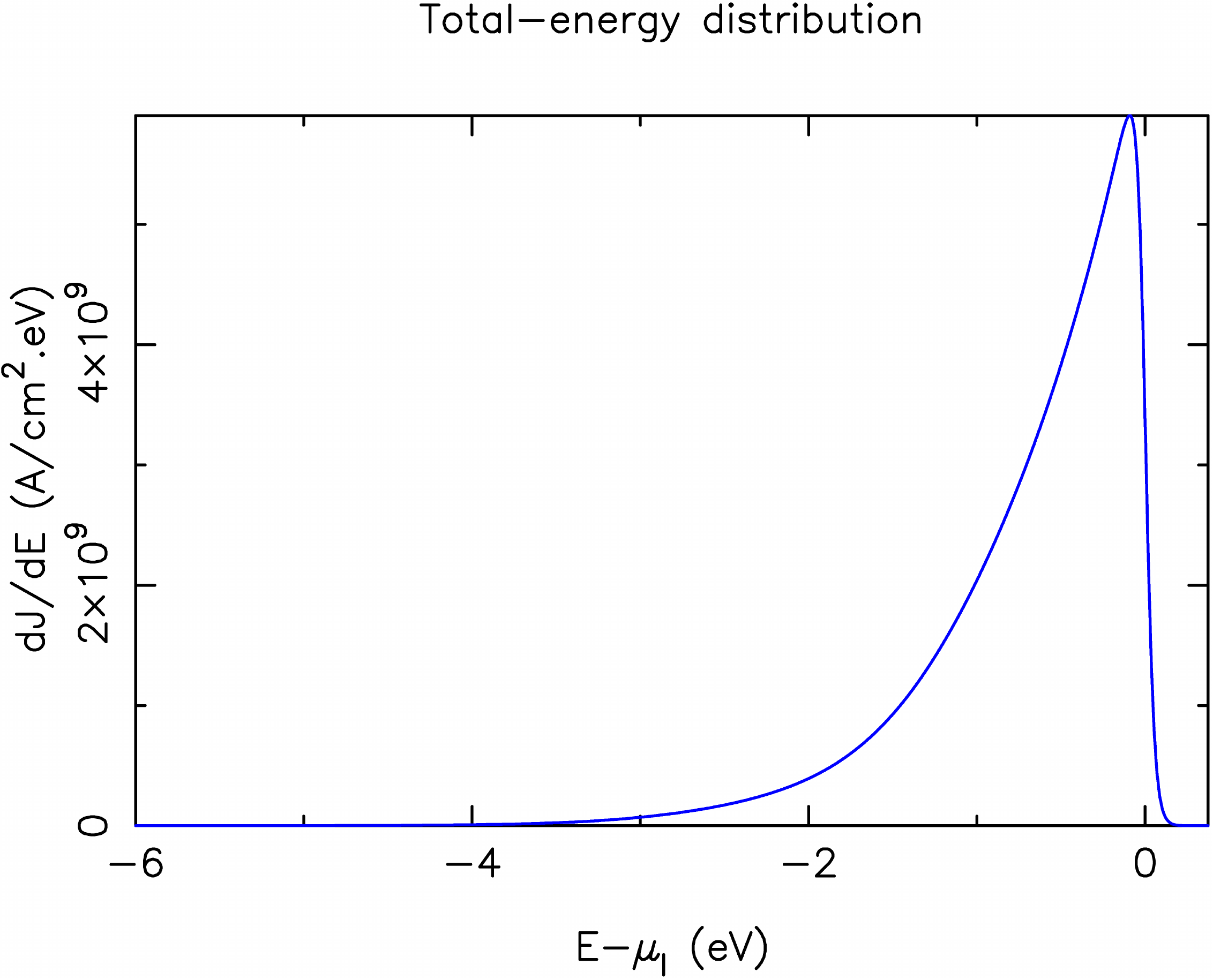} \cr
  \end{tabular}
 \end{center}
 \caption{\label{figure4}Potential energy $V(z)$ (top) and total-energy distribution of the current density $dJ/dE$ (bottom) for an applied voltage ${\sf V}$ of 30 V. $dJ/dE$ is calculated by the transfer-matrix technique. We take for convenience the Fermi level $\mu_{\rm I}$ of the left-side metal as reference for the potential-energy values.}
\end{figure}

We consider for the moment a gap spacing $D$ of 2 nm and three representative values of the applied voltage ${\sf V}$\,: 0.5 V, 5 V and 30 V. The potential-energy distribution $V(z)$ and the total-energy distribution of the current density $dJ/dE$ obtained for these values of the applied voltage are represented in Figs \ref{figure2}, \ref{figure3} and \ref{figure4}. The $dJ/dE$ distributions are calculated by the transfer-matrix technique.

With an applied voltage ${\sf V}$ of 0.5 V (Fig. \ref{figure2}), the Fermi level $\mu_{\rm III}=\mu_{\rm I}-e{\sf V}$ of the right-side metal ("Region III") is 0.5 eV below the Fermi level $\mu_{\rm I}$ of the left-side metal ("Region I"). The
rightwards-moving and leftwards-moving currents in the junction cancel out except in the energy
window between $\mu_{\rm III}$ and $\mu_{\rm I}$ ($\pm$ a few $k_{\rm B}T$,
as a result of the effect of temperature on the electron energy distributions $f_{\rm I}(E)$ and $f_{\rm III}(E)$).
The integrated net current density $J$ that flows from left to right is $1.5\times 10^{-6}$ A/cm$^2$.
We are in the "linear regime" of the $J$-{\sf V} plot. The net current density $J$ depends indeed essentially on the separation between $\mu_{\rm III}$ and $\mu_{\rm I}$, which is equal to $e{\sf V}$.
The mean barrier height $\overline{\phi}$ at the Fermi level is 3.2 eV. Since $e{\sf V} \ll
\overline{\phi}$, Eq. \ref{J_Simmons} will predict a linear $J$-{\sf V} dependence in this regime.

With an applied voltage ${\sf V}$ of 5 V (Fig. \ref{figure3}), the Fermi level $\mu_{\rm III}=\mu_{\rm I}-e{\sf V}$ of the right-side metal is 5 eV below the Fermi level $\mu_{\rm I}$ of the left-side metal. The net current that flows through the junction is essentially determined by the right-flowing current from the left-side metal ("Region I"). The left-flowing current from the right-side metal ("Region III") only contributes for normal energies 5 eV or more below $\mu_{\rm I}$. Its influence on the net current is negligible.
The local current density $J$ that flows from left to right is $6.2$ A/cm$^2$.
The total-energy distribution of the local current density $dJ/dE$ (shown in Fig. \ref{figure3}) is a classical field-emission profile.
The electrons that are emitted by the left-side metal cross the potential-energy barrier in the junction by a
tunneling process. The local current density $J$ increases rapidly with ${\sf V}$. We are in the "field-emission regime" of the $J$-{\sf V} plot. The mean barrier height $\overline{\phi}$ at the Fermi level is 2.6 eV in this case. Since $e{\sf V} >
\overline{\phi}$, Eq. \ref{J_Simmons} will predict a non-linear $J$-{\sf V} dependence.

With an applied voltage ${\sf V}$ of 30 V (Fig. \ref{figure4}), the top $V_{\rm top}$ of the potential-energy barrier drops below the Fermi level $\mu_{\rm I}$ of the left-side metal. All incident
electrons with a normal energy $E_{\rm z}=E-\frac{\hbar^2}{2m}(k_{\rm x}^2+k_{\rm y}^2)>V_{\rm top}$ can actually cross the junction
without tunneling, although quantum-mechanical reflection effects will occur. There is no classical turning point $z_1$ or $z_2$ at the Fermi level $\mu_{\rm I}$ of the left-side metal and Simmons' model for the transmission probability ${\cal D}_{\rm Sim}$ and the local current density $J_{\rm Sim}$ loses any applicability. The mean barrier height $\overline{\phi}$ at the Fermi level can not be calculated in this case since the turning points $z_1$ and $z_2$ are not defined.
We are in the "flyover regime" of the $J$-{\sf V} plot. It is probably interesting for future work to extend
Simmons' theory so that it also applies in this regime. It has been shown by Zhang that
in the flyover regime it is necessary to account for space charge effects.\cite{Zhang_2015}

There is also the possibility that at very high current densities the junction heating will be so great that junction destruction will occur. We are not aware of any work on this effect that is specifically in the context of MVM devices, but for conventional field  electron emitters it is usually thought\cite{Mesyats_1998,Fursey_2005} that heating-related destructive effects will occur for current densities of order $10^7$ to $10^8$ A/cm$^2$ or higher. The situation can become very complicated if in reality there are nanoprotrusions on the emitting surface that cause local field enhancement, and hence local enhancement of the current density, or if heating due to slightly lower current densities can induce the formation and/or growth of nanoprotrusions by means of thermodynamically driven electroformation processes. Detailed examination of these heating-related issues is beyond the scope of the present work.

\begin{figure}[t]
 \begin{center}
  \begin{tabular}{c}
   \includegraphics[height=8.5cm,angle=-90]{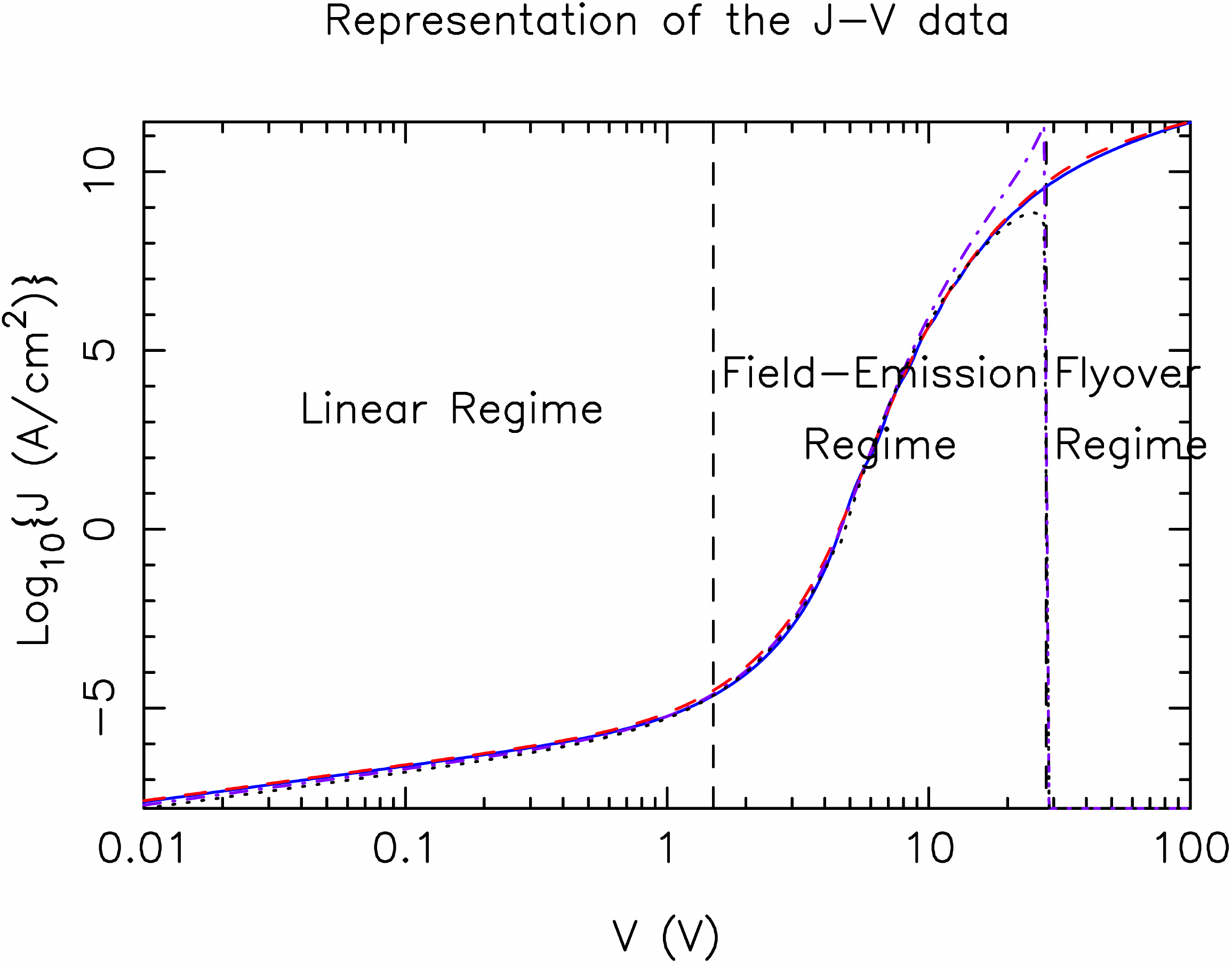}
  \end{tabular}
 \end{center}
 \caption{\label{figure5}$J$-{\sf V} plot for a metal-vacuum-metal junction whose gap spacing $D$ is 2 nm. The four
 curves correspond to $J_{\rm TM}$ (solid), $J_{\rm WKB}$ (dashed), $J_{\rm Sim-num}$ (dot-dashed) and
 $J_{\rm Sim}$ (dotted). These results correspond to a common work function $\Phi$ of 4.5 eV, a Fermi energy
 ${\cal E}_{\rm F}$ of 10 eV and a temperature $T$ of 300 K.}
\end{figure}

The $J$-{\sf V} plot finally obtained for an applied voltage ${\sf V}$ that ranges between 0.01 V and 100 V is represented
in Fig. \ref{figure5}. The figure represents the local current density $J_{\rm TM}$ obtained by the
transfer-matrix technique (Eq. \ref{J_TM} or \ref{J_TM_Ez}; the results are identical), the current density $J_{\rm WKB}$ obtained by a numerical integration of ${\cal D}_{\rm WKB}$ (Eq. \ref{J_WKB_Ez}), the current density $J_{\rm Sim-num}$ obtained by
a numerical integration of ${\cal D}_{\rm Sim}$ (Eq. \ref{J_Simmons_Numeric_Ez}) and the current density
$J_{\rm Sim}$ provided by Simmons' analytical model (Eq. \ref{J_Simmons}). These results
correspond to a gap spacing $D$ of 2 nm. The linear, field-emission and flyover regimes
are clearly indicated. The results provided by the different models turn out
to be in excellent agreement up to a voltage {\sf V} of 10 V. $J_{\rm Sim-num}$ deviates
progressively from the other models beyond this point. The agreement between $J_{\rm TM}$, $J_{\rm WKB}$
and $J_{\rm Sim}$ is remarkable considering the fact the current density varies over 19 orders
of magnitude for the conditions considered. Simmons' analytical model (Eq. \ref{J_Simmons}) turns out to
provide a very good estimate of the current density achieved in the linear and field-emission regimes.
Simmons' analytical model however stops working when Eqs \ref{Simmons_z1z2_2} and \ref{Simmons_Vmoy}
do not provide $\overline{\phi}\geq 0$, which is the case in the flyover
regime (the top of the potential-energy barrier drops indeed below the Fermi level $\mu_{\rm I}$
of the left-side metal and Eq. \ref{T_Simmons} for the transmission probability loses any applicability).

\begin{figure}[t]
 \begin{center}
  \begin{tabular}{c}
   \includegraphics[height=8.5cm,angle=-90]{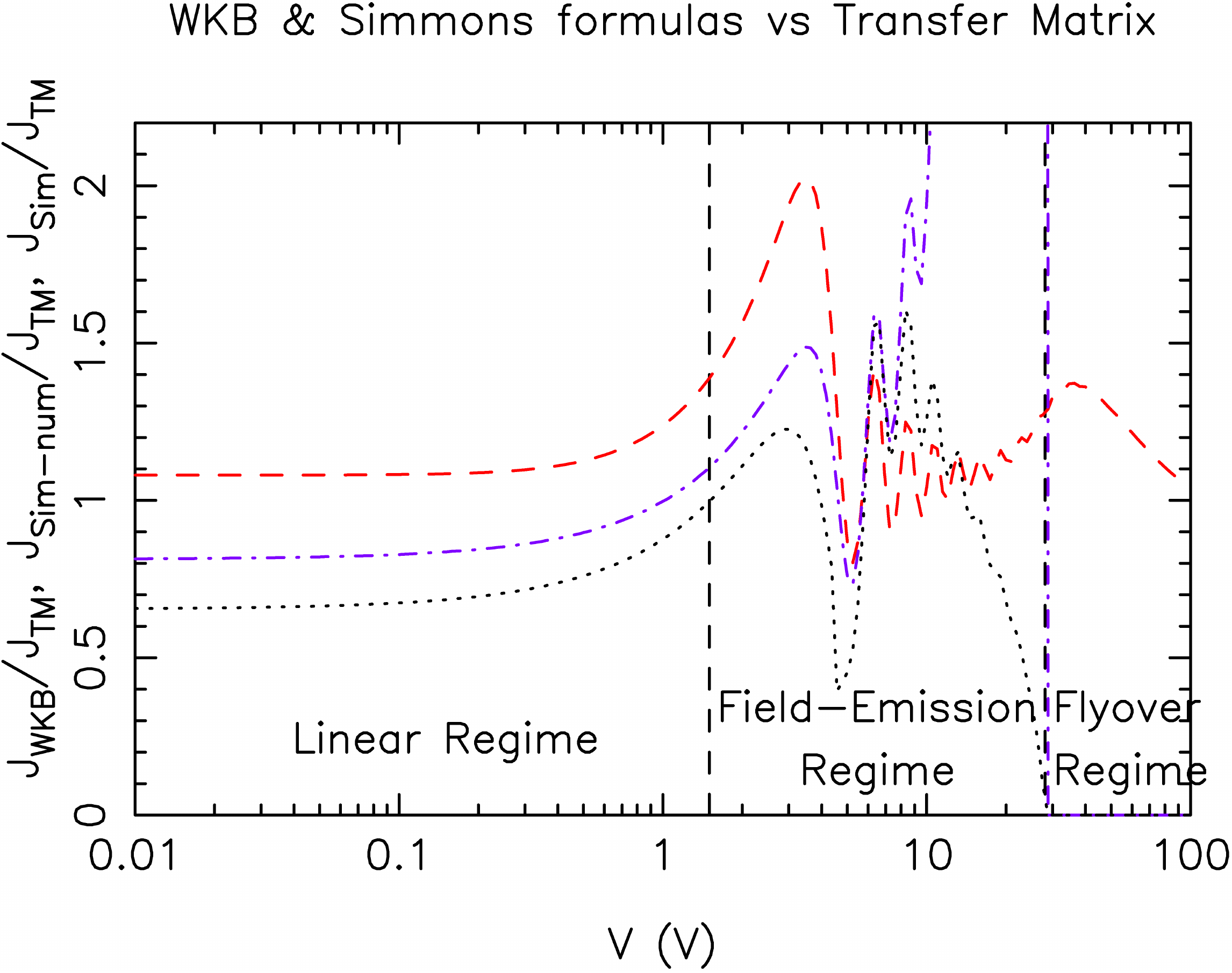}
  \end{tabular}
 \end{center}
 \caption{\label{figure6}Ratio $J_{\rm WKB}/J_{\rm TM}$ (dashed), $J_{\rm Sim-num}/J_{\rm TM}$ (dot-dashed)
  and $J_{\rm Sim}/J_{\rm TM}$ (dotted) for a metal-vacuum-metal junction whose gap spacing $D$ is 2 nm.
  These results correspond to a common work function $\Phi$ of 4.5 eV, a Fermi energy ${\cal E}_{\rm F}$ of 10 eV and a temperature $T$ of 300 K.}
\end{figure}

Figure \ref{figure6} shows more clearly the differences between the different models.
This figure presents the ratio $J_{\rm WKB}/J_{\rm TM}$, $J_{\rm Sim-num}/J_{\rm TM}$ and
$J_{\rm Sim}/J_{\rm TM}$ between the current densities $J_{\rm WKB}$, $J_{\rm Sim-num}$ and $J_{\rm Sim}$
provided by Eqs \ref{J_WKB_Ez}, \ref{J_Simmons_Numeric_Ez} and \ref{J_Simmons} and the transfer-matrix
result $J_{\rm TM}$ (Eq. \ref{J_TM_Ez}). The figure shows that $J_{\rm WKB}$,
$J_{\rm Sim-num}$ and $J_{\rm Sim}$ actually follow the transfer-matrix
result $J_{\rm TM}$ within a factor of the order 0.5-2 up to an applied voltage {\sf V} of 10 V.
The current density $J_{\rm WKB}$ obtained by a numerical integration of ${\cal D}_{\rm WKB}$
with respect to normal energy (Eq. \ref{J_WKB_Ez}) follows in general the transfer-matrix result more closely.
The current densities $J_{\rm Sim}$ derived from Simmons' theory
still provides very decent results. $J_{\rm Sim}$ (Eq. \ref{J_Simmons}) is the analytical expression derived by Simmons (main focus of this article). $J_{\rm WKB}$ and $J_{\rm Sim-num}$ require a numerical evaluation of the transmission probability
(by Eq. \ref{T_WKB} or \ref{T_Simmons}) and a numerical integration of this transmission
probability with respect to normal energy to finally obtain the current density. They are presented only for comparison.
We note that $J_{\rm WKB}$ tends here to overestimate the local current densities.
This behavior was already observed with the Schottky-Nordheim barrier that is relevant to
field electron emission from a flat metal, when considering normal energies in the vicinity
of the Fermi level of a metal whose physical parameters are the same as those considered at this point
($\Phi$=4.5 eV and ${\cal E}_{\rm F}$=10 eV).\cite{Mayer_2010_JPCM,Mayer_2010_JVSTB1} As shown in Ref.
\citenum{Mayer_2010_JVSTB2}, underestimation of the local current densities by the simple WKB approximation
is also possible for smaller values of ${\cal E}_{\rm F}$.
We note finally that $J_{\rm Sim-num}$ and $J_{\rm Sim}$ provide close results up to
an applied voltage ${\sf V}$ of 10 V. This proves that the approximations that lead to $J_{\rm Sim}$
are reasonable up to this point.
$J_{\rm Sim-num}$, which is based on a numerical integration of ${\cal D}_{\rm Sim}$, starts then over-estimating
the current density. Simmons' mean-barrier approximation is actually a poor model of the transmission
probability when the potential-energy barrier becomes too small (we can indeed have $E_{\rm z}-\mu_{\rm I}>\overline \phi$
for values of $E_{\rm z}$ that have a non-negligible $\Delta {\cal N}(E_{\rm z})$, while in reality $E_{\rm z}-\mu_{\rm I}<\phi(z)$
in the potential-energy barrier). Simmons' analytical expression for the local
current density ($J_{\rm Sim}$ by Eq. \ref{J_Simmons}) appears to be more robust in these conditions.
$J_{\rm Sim-num}$ and $J_{\rm Sim}$ cannot be applied in the flyover regime.

\begin{table}[t]
 \begin{center}
 \begin{tabular}{|c||c|c|c|c|}
  \hline
   & \multicolumn{4}{c|}{$D$=0.5 nm} \cr
  \hline
  $\Phi$ (eV) & ${\sf V}$=0.01 V & ${\sf V}$=0.1 V & ${\sf V}$=1 V & ${\sf V}$=10 V \cr
  \hline
  1.5 & /     & /     & /     & / \cr
  2.0 & /     & /     & /     & / \cr
  2.5 & /     & /     & /     & / \cr
  3.0 & /     & /     & /     & / \cr
  3.5 & 0.362 & 0.367 & 0.327 & / \cr
  4.0 & 0.470 & 0.478 & 0.558 & / \cr
  4.5 & 0.481 & 0.489 & 0.587 & / \cr
  5.0 & 0.462 & 0.470 & 0.562 & / \cr
  \hline
  \hline
   & \multicolumn{4}{c|}{$D$=1 nm} \cr
  \hline
  $\Phi$ (eV) & ${\sf V}$=0.01 V & ${\sf V}$=0.1 V & ${\sf V}$=1 V & ${\sf V}$=10 V \cr
  \hline
  1.5 & 0.872 & 0.871 & /     & /     \cr
  2.0 & 1.811 & 1.898 & 2.605 & /     \cr
  2.5 & 1.562 & 1.630 & 2.550 & /     \cr
  3.0 & 1.265 & 1.312 & 1.969 & /     \cr
  3.5 & 1.029 & 1.062 & 1.511 & /     \cr
  4.0 & 0.852 & 0.876 & 1.189 & 0.088 \cr
  4.5 & 0.721 & 0.739 & 0.964 & 1.494 \cr
  5.0 & 0.622 & 0.635 & 0.802 & 1.993 \cr
  \hline
  \hline
   & \multicolumn{4}{c|}{$D$=2 nm} \cr
  \hline
  $\Phi$ (eV) & ${\sf V}$=0.01 V & ${\sf V}$=0.1 V & ${\sf V}$=1 V & ${\sf V}$=10 V \cr
  \hline
  1.5 & 2.781 & 3.056 & 3.670 & /     \cr
  2.0 & 2.137 & 2.297 & 3.604 & /     \cr
  2.5 & 1.594 & 1.687 & 2.633 & /     \cr
  3.0 & 1.218 & 1.275 & 1.893 & 0.961 \cr
  3.5 & 0.962 & 0.999 & 1.409 & 1.205 \cr
  4.0 & 0.784 & 0.809 & 1.092 & 1.482 \cr
  4.5 & 0.656 & 0.674 & 0.877 & 1.259 \cr
  5.0 & 0.563 & 0.576 & 0.726 & 1.097 \cr
  \hline
  \hline
   & \multicolumn{4}{c|}{$D$=5 nm} \cr
  \hline
  $\Phi$ (eV) & ${\sf V}$=0.01 V & ${\sf V}$=0.1 V & ${\sf V}$=1 V & ${\sf V}$=10 V \cr
  \hline
  1.5 & 1.328 & 1.411 & 0.391 & /     \cr
  2.0 & 1.384 & 1.500 & 1.349 & 0.683 \cr
  2.5 & 1.080 & 1.150 & 1.362 & 0.847 \cr
  3.0 & 0.851 & 0.895 & 1.118 & 0.814 \cr
  3.5 & 0.689 & 0.717 & 0.897 & 0.700 \cr
  4.0 & 0.572 & 0.592 & 0.731 & 0.565 \cr
  4.5 & 0.487 & 0.501 & 0.608 & 0.434 \cr
  5.0 & 0.423 & 0.434 & 0.518 & 0.310 \cr
  \hline
 \end{tabular}
 \caption{\label{table1}Ratio $J_{\rm Sim}/J_{\rm TM}$ between the local current density $J_{\rm Sim}$
  provided by Simmons' analytical model and the current density $J_{\rm TM}$ provided by the transfer-matrix technique,
  for different values of the gap spacing $D$, the common metal work function $\Phi$ and the applied voltage ${\sf V}$.
  The Fermi energy ${\cal E}_{\rm F}$ is 10 eV and the temperature $T$ is 300 K.}
 \end{center}
\end{table}

We finally provide in Table \ref{table1} a more systematic study of the ratio $J_{\rm Sim}/J_{\rm TM}$ between the current density $J_{\rm Sim}$ provided by Simmons' analytical model (Eq. \ref{J_Simmons}) and the current density $J_{\rm TM}$ provided by the transfer-matrix
technique (Eq. \ref{J_TM_Ez}). These $J_{\rm Sim}/J_{\rm TM}$ ratios are calculated for different values of the gap spacing $D$, work function $\Phi$ and applied voltage ${\sf V}$. The values considered for $D$ (0.5, 1, 2 and 5 nm), $\Phi$ (1.5, 2,... 5 eV) and ${\sf V}$ (0.01, 0.1, 1 and 10 V) are of practical interest when applying Simmons' theory for the current density in metal-vacuum-metal junctions.
The results show that Simmons' analytical expression for the local current density actually provides results that are in a good agreement with those provided by the transfer-matrix technique. The factor $J_{\rm Sim}/J_{\rm TM}$ that expresses the difference between the two models is of the order of 0.3-3.7 in most cases. Simmons' model obviously loses its applicability when Eq. \ref{Simmons_Vmoy} for $\overline{\phi}$ predicts a mean barrier height at the left-side Fermi level $\overline{\phi}<0$. In conditions for which $\overline{\phi}\geq0$, Simmons' analytical expression (Eq. \ref{J_Simmons}) turns out to provide decent estimations of the current density $J$ that flows in the metal-vacuum-metal junction considered in this work. This justifies the use of Simmons' model for these systems.

It has been assumed in this modelling paper that both electrodes are smooth, flat and planar. This may not be an adequate modelling approximation and it may be that in some real devices the electrostatic field near the emitting electrode varies somewhat across the electrode surface. In such cases, the "real average current density"
is probably better expressed as $J_{\rm av} = \alpha_{\rm n}\ J_{\rm local}$, where $J_{\rm local}$ is the local current density at a typical hot spot and the parameter $\alpha_{\rm n}$ (called here the "notional area efficiency") is a measure of the apparent fraction of the electrode area that is contributing significantly to the current flow. However, there is no good present knowledge of the values of either of these quantities.
It is also necessary to be aware that smooth-surface conceptual models disregard the existence of atoms and do not attempt to evaluate the role that atomic-level wave-functions play in the physics of tunneling.
In the context of field electron emission,\cite{Fowler_1939,Sommerfeld_1964,Ziman_1964} it is known that these smooth-surface models are unrealistic and that the neglect of atomic-level effects creates uncertainty over the predictions of the smooth-surface models. At present, it is considered that the derivation of accurate atomic-level theory is a very difficult problem, so reliable assessment of the error in the smooth-surface models is not possible at present.
However, in the context of field electron emission, our present guess is that the smooth-surface models may over-predict
by a factor of up to 100 or more, or under-predict by a factor of up to 10 or more.
Recent results obtained by Lepetit are consistent with these estimations.\cite{Lepetit_2017}
Uncertainties of this general kind will also apply to the Simmons results and to the results derived in this paper.

\section{\label{section5}Conclusions}

We used a transfer-matrix technique to test the consistency with which Simmons’ analytical model actually predicts the local current density $J$ that flows in flat metal-vacuum-metal junctions. Simmons' analytical model relies on a mean-barrier approximation for the transmission probability. This enables the derivation of an analytical expression for the current density. In Simmons' original
papers, there is a missing factor $1/2$ in the image potential energy. This factor was included for correction in our presentation
of Simmons' theory. We then compared the current density $J_{\rm Sim}$ provided by this analytical model with the
current density $J_{\rm TM}$ provided by a transfer-matrix technique. We also considered the current densities provided by a numerical
integration of the transmission probability obtained with the WKB approximation and Simmons' mean-barrier approximation.
The comparison between these different models shows that Simmons' analytical model for the current density provides results that are
in good agreement with an exact solution of Schr\"odinger's equation for a range of conditions of practical interest. The ratio
$J_{\rm Sim}/J_{\rm TM}$ used to measure the accuracy of Simmons' model takes values of the order of 0.3-3.7 in most
cases, for the conditions considered in this work. Simmons' model can obviously only be used when the mean-barrier height at the
Fermi level $\overline\phi$ is positive. This corresponds to the linear and field-emission regimes of $J$-{\sf V} plots.
Future work may extend the range of conditions considered for this numerical testing of Simmons' model
and seek at establishing a correction factor to use with Simmons' equation in order to get an exact result.

\acknowledgments

A.M. is funded by the Fund for Scientific Research (F.R.S.-FNRS) of
Belgium. He is member of NaXys, Namur Institute for Complex Systems,
University of Namur, Belgium. This research used resources of the
``Plateforme Technologique de Calcul Intensif (PTCI)''
(http://www.ptci.unamur.be) located at the University of Namur,
Belgium, which is supported by the F.R.S.-FNRS under the convention
No. 2.5020.11. The PTCI is member of the ``Consortium des
Equipements de Calcul Intensif (CECI)'' (http://www.ceci-hpc.be).


\begin{thebibliography}{99}
 \bibitem{Jeffreys_1925} H. Jeffreys, "On Certain Approximate Solutions of Linear Differential Equations of the Second Order," Proc. London Math. Soc. {\bf s2-23}, 428--436 (1925).
%
 \bibitem{Wentzel_1926} G. Wentzel, "Eine Verallgemeinerung der Quantenbedingungen {f\"ur} die Zwecke der Wellenmechanik," Z. Phys. {\bf 38}, 518--529 (1926).
%
 \bibitem{Kramers_1926} H.A. Kramers, "Wellenmechanik und halbzahlige Quantisierung," Z. Phys. {\bf 39}, 828--840 (1926).
%
 \bibitem{Brillouin_1926} L. Brillouin, "La m\'ecanique ondulatoire de {Schr\"odinger}: une m\'ethode g\'en\'erale de r\'esolution par approximations successives," Compt. Rend. {\bf 183}, 24--26 (1926).
%
 \bibitem{Fowler_Nordheim_1928} R.H. Fowler and L. Nordheim, "Electron emission in intense electric fields," Proc. R. Soc. London Ser. A {\bf 119}, 173--181 (1928).
%
 \bibitem{Murphy_Good_1956} E.L. Murphy and R.H. Good, "Thermionic Emission, Field Emission, and the Transition Region," Phys. Rev. {\bf 102}, 1464--1473 (1956).
%
 \bibitem{Good_Muller_1956} R.H. Good and E.W. {M\"uller}, "Field Emission" in {\it Handbuch der Physik} (Springer Verlag, Berlin, 1956), pp 176--231.
%
 \bibitem{Young_1959} R.D. Young, "Theoretical Total-Energy Distribution of Field-Emitted Electrons," Phys. Rev. {\bf 113}, 110--114 (1959).
%
 \bibitem{Forbes_JAP_2008} R.G. Forbes, "On the need for a tunneling pre-factor in Fowler–Nordheim tunneling theory," J. Appl. Phys. {\bf 103}, 114911 (2008).
%
 \bibitem{Forbes_JVSTB_2008} R.G. Forbes, "Physics of generalized Fowler-Nordheim-type equations," J. Vac. Sci. Technol. B {\bf 26}, 788--793 (2008).
%
 \bibitem{Mayer_2010_JPCM} A. Mayer, "A comparative study of the electron transmission through one-dimensional barriers relevant to field-emission problems," J. Phys. Condens. Matter {\bf 22}, 175007 (2010).
%
 \bibitem{Mayer_2010_JVSTB1} A. Mayer, "Numerical testing of the Fowler-Nordheim equation for the electronic field emission from a flat metal and proposition for an improved equation," J. Vac. Sci. Technol. B {\bf 28}, 758--762 (2010).
%
 \bibitem{Mayer_2010_JVSTB2} A. Mayer, "Exact solutions for the field electron emission achieved from a flat metal using the standard Fowler-Nordheim equation with a correction factor that accounts for the electric field, the work function and the Fermi energy of the emitter," J. Vac. Sci. Technol. B {\bf 29}, 021803 (2010).
%
 \bibitem{Hagmann_1995} M.J. Hagmann, "Efficient numerical methods for solving the Schr\"odinger equation with a potential varying sinusoidally with time," Int. J. Quantum Chem. {\bf 29}, 289--295 (1995).
%
 \bibitem{Forbes_2007} R.G. Forbes and J.H.B. Deane, "Reformulation of the standard theory of Fowler-Nordheim tunnelling and cold field electron emission," Proc. R. Soc. A {\bf 463}, 2907--2927 (2007).
%
 \bibitem{Simmons_1963_JAP1} J.G. Simmons, "Generalized formula for the electric tunnel effect between similar electrodes separated by a thin insulating film," J. Appl. Phys. {\bf 34}, 1793--1803 (1963).
%
 \bibitem{Simmons_1963_JAP2} J.G. Simmons, "Electric tunnel effect between dissimilar electrodes separated by a thin insulating film," J. Appl. Phys. {\bf 34}, 2581--2590 (1963).
%
 \bibitem{Simmons_1964_JAP1} J.G. Simmons, "Potential barriers and emission-limited current flow between closely spaced parallel metal electrodes," J. Appl. Phys. {\bf 35}, 2472--2481 (1964).
%
 \bibitem{Simmons_1964_JAP2} J.G. Simmons, "Generalized thermal {J-V} characteristic for the electric tunnel effect," J. Appl. Phys. {\bf 35}, 2655--2658 (1964).
%
 \bibitem{Matthews_2018} N. Matthews, M.J. Hagmann and A. Mayer, "Comment: Generalized formula for the electric tunnel effect between similar electrodes separated by a thin insulating film," J. Appl. Phys. {\bf 123}, 136101 (2018).
%
 \bibitem{Miskovsky_1982} N.M. Miskovsky, P.H. Cutler, T.E. Feuchtwang and A.A. Lucas, "The multiple-image interactions and the mean-barrier approximation in MM and MVM tunneling junctions," Appl. Phys. A {\bf 27}, 139--147 (1982).
%
 \bibitem{Mayer_PRE_1999} A. Mayer and J.-P. Vigneron, "Accuracy-control techniques applied to stable transfer-matrix computations," Phys. Rev. E {\bf 59}, 4659--4666 (1999).
%
 \bibitem{Footnote_2} One can check indeed that $\Psi_{i,j}^{+} = \hat\Psi_{i,j}^{+}\  [T_{i,j}^{++}]^{-1}$ and $\Psi_{i,j}^{-} = \hat\Psi_{i,j}^{-} - \hat\Psi_{i,j}^{+}\ [T_{i,j}^{++}]^{-1} T_{i,j}^{+-}$.
%
 \bibitem{Mayer_PRB_1997} A. Mayer and J.-P. Vigneron, "Real-space formulation of the quantum-mechanical elastic diffusion under n-fold axially symmetric forces," Phys. Rev. B {\bf 56}, 12599--12607 (1997).
%
 \bibitem{Mayer_PRB_2008} A. Mayer, M.S. Chung, B.L. Weiss, N.M. Miskovsky and P.H. Cutler, "Three-dimensional analysis of the rectifying properties of geometrically asymmetric metal-vacuum-metal junctions treated as an oscillating barrier," Phys. Rev. B {\bf 78}, 205404 (2008).
%
 \bibitem{Mayer_JVSTB_2012} A. Mayer, M.S. Chung, P.B. Lerner, B.L. Weiss, N.M. Miskovsky and P.H. Cutler, "Analysis of the efficiency with which geometrically asymmetric metal-vacuum-metal junctions can be used for the rectification of infrared and optical radiations", J. Vac. Sci. Technol. B {\bf 30}, 31802 (2012).
%
 \bibitem{Footnote_3} The use of factors of the form $f_{\rm I}(E)[1-f_{\rm III}(E)]$ and $f_{\rm III}(E)[1-f_{\rm I}(E)]$ in the two terms of the current density will provide identical results when the applied voltage ${\sf V}$ is static as in this work. For oscillating voltages, the expression \ref{J_TM} must however be used and we consider it therefore as fundamentally more correct.
%
 \bibitem{Footnote_1} For the transfer-matrix calculations, it is the expression \ref{Potential_Energy} that is actually used for $V(z)$. The fact $V_{\rm image}(z)$ tends to $-\infty$ when $z\to 0$ or $z\to D$ causes convergence issues when solving Schr\"odinger's equation by a transfer-matrix approach. The physical reason is related to the existence of bound states in this potential energy. These bound states would be filled in the real device. One can actually question the validity of the image interaction when we are at a few Angstr\"oms only to the surface of a metal. A solution to this issue is to cut $V(z)$ at $V_{\rm I}$ when $z\to 0$ and at $V_{\rm III}$ when $z\to D$. This provides the barrier depicted in Fig. \ref{figure1} in which there is
     no singularity in the potential energy when crossing the surface of each metal.
%
 \bibitem{Hrach_1968} R. Hrach, "A contribution to the temperature dependence of the tunnel current of metal-dielectric-metal structures," Czech. J. Phys. B {\bf 18}, 402--418 (1968).
%
 \bibitem{Zhang_2015} P. Zhang, "Scaling for quantum tunneling current in nano- and subnano-scale plasmonic junctions," Sci. Rep. {\bf 5}, 09826 (2015).
%
 \bibitem{Mesyats_1998} G.A. Mesyats, Explosive Electron Emission (URO Press, Ekaterinburg, 1998).
%
 \bibitem{Fursey_2005} G. Fursey, Field Emission in Vacuum Microelectronics (Kluwer/Plenum, New York, 2005).
%
 \bibitem{Fowler_1939} R.H. Fowler and E.A. Guggenheim, {\it Statistical Thermodynamics} 2nd edition (Cambridge Univ. Press, London, 1949).
%
 \bibitem{Sommerfeld_1964} A. Sommerfeld, {\it Thermodynamics and Statistical Mechanics} (Academic Press, New York, 1964).
%
 \bibitem{Ziman_1964} J.M. Ziman, {\it Principles of the Theory of Solids} (Cambridge Univ. Press, London, 1964).
%
 \bibitem{Lepetit_2017} B. Lepetit, "Electronic field emission models beyond the Fowler-Nordheim one," J. Appl. Phys. {\bf 122}, 215105 (2017).
%
\end{thebibliography}
\end{document}